\documentclass[a4paper,12pt]{article}
\usepackage{soul}
\usepackage[usenames,dvipsnames]{color}
\usepackage{jheppub}
\usepackage{amsmath, amssymb, slashed, epsf, color, graphicx, latexsym}
\usepackage{epsfig}
\usepackage{graphics}

\begin{document}

\title{Dual Theory for maximally $\mathcal{N}$ extended flat Supergravity}

\author[a]{Nabamita Banerjee,}
\author[b]{Arindam Bhattacharjee,}
\author[a]{Surajit Biswas,}
\author[c]{and Turmoli Neogi}
\affiliation[a]{Indian Institute of Science Education and Research Bhopal,
	Bhopal Bypass, Bhopal 462066, India.}
\affiliation[b]{Indian Institute of Science Education and Research, Pune 411008, India}
\affiliation[c]{Universit\'e Libre de Bruxelles and International Solvay Institutes,\\
            Campus Plain - CP 231, B-1050 Bruxelles, Belgium}
	
\emailAdd{nabamita@iiserb.ac.in}
\emailAdd{arindam.bhattacharjee@students.iiserpune.ac.in}
\emailAdd{surajit18@iiserb.ac.in}
\emailAdd{turmoli.neogi@ulb.be}

\abstract{Maximally $\mathcal{N}$ extended $2+1$ dimensional flat Supergravity theories exist for a class of super-Poincar\'{e} algebras for arbitrary $\mathcal{N}$. They have rich asymptotic structures and contain all interesting topological supergravity solutions in presence of non-trivial holonomy. For the asymptotic symmetry algebra being a suitable flat limit of the superconformal algebra, which is an infinite dimensional extension of two copies of $Osp(\mathcal{N}|2;R)$ group, we have found the $1+1$ dimensional chiral WZW model as the dual quantum field theory that describes the dynamics of these solutions. In the Hamiltonian analysis, the reduced phase space resembles a flat super Liouville theory.}

\maketitle
\section{Introduction and Summary}
Gravity in $2+1$ space time dimensions is very interesting. In $2+1$ dimensions it becomes a renormalizable theory and hence one can ask interesting questions about its quantum aspects (\cite{Carlip:1995zj} and the references therein). Another speciality of $2+1$ dimensions is that gravity becomes topological in nature and hence does not have any dynamical on-shell bulk degrees of freedom. All solutions are locally equivalent to the Minkowski spacetime geometry\footnote{In presence of a cosmological constant, depending on the sign, one gets locally a dS or AdS geometry}. Thus we do not have graviton or gravitational waves in $2+1$ dimensions. Nevertheless, whenever the embedding spacetime has noncontractible cycles, nontrivial boundary topologies arise and one ends up getting boundary gravitons. These boundary degrees of freedom play the most important role in $2+1$ gravity theories. Since the dynamics lies on the boundary, it is natural to think that gravity in $2+1$ dimensions must have a dual description in terms of a $1+1$ dimensional theory living on the boundary of the manifold.  
The duality is best understood in the Chern-Simons (CS) formulation of gravity \cite{Witten:1988hc, Achucarro:1986uwr}. In $2+1$ dimensions, both gravity and CS theory are topological and they are equivalent descriptions of each other for the CS gauge group $ISO(2,1)$. In past this equivalence has been mostly explored for the case of asymptotically AdS gravities \cite{Witten:1983ar}. By now for all three interesting asymptotic structures namely flat, Anti de-sitter (AdS) and de-sitter (dS), the dual boundary field theory for ordinary Einstein-Hilbert gravity has been found \cite{Witten:1988hf, Moore:1989yh, Elitzur:1989nr, Compere:2014cna}. The duality is classical in nature, where the asymptotic symmetries of the gravity theory matches with the symmetry of the boundary dual field theory. In this paper we are interested in asymptotically flat gravity theories, where the asymptotic symmetry in known as the three dimensional BMS symmetry.

  The supersymmetric extensions of BMS$_3$ algebras were derived in \cite{Barnich:2014cwa} and later they were shown to be the asymptotic symmetries for $2+1$ dimensional supergravity theories by explicit gravity computations. Extended Supersymmetry brings more interesting features to these gravity theories. One gets an infinite set of regular 3D topological gravity solutions with non trivial holonomy that makes them locally different from the ordinary Minkowski solution. Such purely bosonic solutions, while setting the dynamical boundary graviton modes to constants, were constructed in \cite{Banerjee:2018hbl}. It was found that, in presence of nontrivial $R-symmetry$ in the bulk, these topological bosonic solutions range over non-supersymmetric, null orbifold (preserving half supersymmetry), conical defect (preserving full supersymmetry) to conical surplus types of solutions.
The dual boundary field theories are also known for minimal and higher $\mathcal{N}=2,4$ supergravity theories for asymptotically flat and AdS geometries \cite{Coussaert:1995zp,Floreanini:1987as,article,SONNENSCHEIN1988752,PhysRevLett.62.1817,Chu:1991pn,Caneschi:1996sr,Henneaux:1999ib,Valcarcel:2018kwd, Barnich:2015sca, Barnich:2012rz,Banerjee:2015kcx}. The construction trivially extends to $\mathcal{N}=8$ supergravity theory as well. For $\mathcal{N}=8$ supergravity we consider the corresponding R-symmetry group to be $SU(2)$ and the corresponding generators belong to its fundamental representation. Recently the dual theory for minimal supersymmetric gravity theory with dS asymptotic geometry has also been constructed \cite{Bhattacharjee:2020jpq}. In all these cases, the dual descriptions turn out to be a gauged chiral Wess-Zumino-Witten model with $SL(2,R)$ or its appropriate supersymmetric extensions as the gauge group. 

In this paper we look for the dual description of asymptotically flat $2+1$ dimensional maximally $\mathcal{N}$ extended supergravity theory, with arbitrary $\mathcal{N}$, i.e. the number of supercharges for these theories is kept arbitrary. The bulk symmetry algebras are just super-Poincar\'{e} algebras with arbitrary $\mathcal{N}$. The asymptotic symmetry for this system and the generic bosonic gravitational solutions were studied in \cite{Banerjee:2018hbl}. The asymptotic symmetry algebra turns out to be a suitable flat limit of the superconformal algebra (SCA), that we name as $\mathcal{N}$ extended super-BMS$_3$ algebra. It was found that the asymptotic symmetry forms a non-linear algebra as it closes with non-linear quadratic terms involving the generators. A large number of such SCAs of 
interest were considered in \cite{Bina_1997,Henneaux:1999ib} and similar interesting feature was also seen in the corresponding supergravity with asymptotically AdS geometry and was extensively presented in \cite{Henneaux:1999ib}. Thus, in this case, even the zero mode sector of the asymptotic symmetry algebra is centrally extended compared to the bulk super-Poincar\'{e} algebra. In the present paper we find the $1+1$ dimensional dual of such asymptotically flat $\mathcal{N}$ supergravity theories. Our result states that a chiral Wess-Zumino-Witten (WZW) model type dual for such $\mathcal{N}$ supergravity theories is possible to achieve only when its $\mathcal{N}$ extended super-BMS$_3$ asymptotic symmetry algebra is a suitable flat limit of SCA which is an infinite dimensional extension of two copies of  $Osp(N|2;R)$. The WZW model has the super-Poincar\'{e} algebra as its gauge group. An appropriate Hamiltonian phase space analysis is performed and the reduced phase space is found to be a supersymmetric extension of the Liouville theory in its flat limit.

The paper is organised as follows : In \ref{sec2} we present the bulk maximally $\mathcal{N}$ extended Super-Poincar\'{e} algebra which is the basis of our construction. The $2+1$ dimensional supergravity theory that we are interested in has these superalgebras as its bulk symmetry. We also construct the supertrace elements of the bulk algebra. In section \ref{sec3} we present the supergravity theory in Chern-Simons formulation and find its gauge fixed classical solutions. This gives us the non-trivial residual boundary modes that can not be fixed classically. Next in section \ref{sec4}, we present the dual boundary chiral Wess-Zumino-Witten (WZW) model that governs the dynamics of the boundary modes. The WZW model has the $\mathcal{N}$ extended Super-Poincar\'{e} group as its gauge group. We also present the gauge version of the WZW model. We further find the reduced phase space description of the theory in section \ref{sec5} and find that the phase space is governed by a flat limit of super Liouville theory. The paper is concluded in section \ref{sec6}. There are two appendices \ref{App1} and \ref{App2} that summarize our notations, conventions and useful identities.

\section{The 3D maximally $\mathcal{N}$ extended super-Poincar\'{e} Algebra}\label{sec2}

Our aim is to find the two-dimensional theory dual to three dimensional maximally $\mathcal{N}$-extended asymptotically flat supergravity, for arbitrary $\mathcal{N}$.
A list of such interesting algebras were presented in \cite{Banerjee:2016nio}. For this purpose, we consider the most generic $\mathcal{N}$-extended super-Poincar\'{e} algebra with following non zero elements:
\begin{align}
&[J_a, J_b] = \epsilon_{abc}J^c \qquad [J_a, P_b] = \epsilon_{abc}P^c \nonumber \\
&[T^{\bar{a}}, T^{\bar{b}}] = f^{\bar{a}\bar{b}}_{\,\,\,\,\,\,\bar{c}} T^{\bar{c}} \qquad [T^{\bar{a}}, Z^{\bar{b}}] = f^{\bar{a}\bar{b}}_{\,\,\,\,\,\,\bar{c}} Z^{\bar{c}} \nonumber \\
&[J_a, Q^{(1,2),\alpha}_{\hat{p}}] = \frac{1}{2} (\Gamma^a)^{\hat{q}}_{\,\,\,\hat{p}} Q^{(1,2),\alpha}_{\hat{q}} \quad [T^{\bar{a}}, Q^{1,\alpha}_{\hat{p}}] = (\lambda^{\bar{a}})^{\alpha}_{\,\,\,\beta} Q^{1,\beta}_{\hat{p}}\quad [T^{\bar{a}}, Q^{2,\alpha}_{\hat{p}}] = -(\lambda^{\bar{a}})^{\alpha}_{\,\,\,\beta} Q^{2,\beta}_{\hat{p}} \nonumber \\
&\left\lbrace  Q^{1,\alpha}_{\hat{p}}, Q^{1,\beta}_{\hat{q}}\right\rbrace = -\frac{1}{2} \eta^{\alpha\beta} (C\Gamma^a)_{\hat{p}\hat{q}} P_a + \frac{1}{6\hat{\alpha}} C_{\hat{p}\hat{q}} (\lambda^{\bar{a}})^{\alpha\beta} Z_{\bar{a}} \label{eq:superPoincare}  \\
&\left\lbrace  Q^{2,\alpha}_{\hat{p}}, Q^{2,\beta}_{\hat{q}}\right\rbrace = -\frac{1}{2} \eta^{\alpha\beta} (C\Gamma^a)_{\hat{p}\hat{q}} P_a - \frac{1}{6\hat{\alpha}} C_{\hat{p}\hat{q}} (\lambda^{\bar{a}})^{\alpha\beta} Z_{\bar{a}}. \nonumber
\end{align}
Our $2+1$ dimensional asymptotically flat gravity theory in invariant under the above bulk symmetry algebra.
Here the first line is the bosonic spacetime algebra. The indices $a,b$ runs over values $\{1,2,3 \}$. The second line is the automorphism algebra, the barred indices $\bar{a}, \bar{b}..$ run over $\{1,2,...D \}$ where $D$ is the dimension of the internal algebra. The indices $\alpha, \beta$ go from $1,...,d$ and denote the amount of supersymmetry $\mathcal{N}$. Here $d$ is the dimension of the real representation $\rho$ of the automprhism, under which the fermions transform. Indices $\hat{p}, \hat{q}..$ run over $\{\frac{1}{2},-\frac{1}{2} \}$. The metric in $(\bar a, \bar b)$ space is $\delta_{\bar a \bar b}$ and in the internal algebra space $(\alpha\beta)$ is $\eta_{\alpha\beta}$\footnote{it is not to be confused with the Minkowski metric.}. The constant $\hat{\alpha}$ is related to the representation of the automorphism algebra by $\hat{\alpha} = \frac{C_{\rho}}{3(d-1)}$, where $C_{\rho}$ is the second Casimir of the internal algebra at representation $\rho$. A specific case when the superalgebra which is particularly interesting for our study is when the internal symmetry group is $so(\mathcal{N})$. In this case, $D$ and $d$ are realted as $D=\frac{d(d-1)}{2}$. Our notations and conventions are further stated in Appendix \ref{App1}. \\
Let us make a few comments about the internal algebra \cite{Bina_1997} in general. The generators $(\lambda^{\bar{a}})^{\alpha\beta}$ of representation $\rho$ of the internal algebra are orthogonal and anti-symmetric in their indices. Thus, the dimension of the internal algebra $d(=\mathcal{N})$ is even.  The prime point is that the anti-commutator of $(\lambda^{\bar{a}})$'s  are not constrained and is kept arbitrary in the algebra. Nevertheless the algebra is highly constrained by the Jacobi identities. Various important properties of the generators are listed in Appendix \ref{App2} .
~~\\

\subsection*{The Supertrace Elements}
In the context of the CS formulation of gravity theory, an symmetry algebra is interesting if it admits a non-degenerate invariant bi-linear (quadratic Casimir form). (Such construction of supertrace elements were considered in \cite{Howe:1995zm,Giacomini:2006dr}). This condition works for all $\mathcal{N}$ extended superalgebras listed in \cite{Banerjee:2016nio}. To find the invariant non-degenerate bi-linear exists, we construct the most generic quadratic combination of the generators with arbitrary coefficients. For it to be a Casimir, this quadratic form must commute with every generator of the algebra. These put conditions on the arbitrary coefficients. If these equations can be solved consistently, we get a quadratic Casimir in terms of the coefficients determined by these conditions. Then the supertrace elements come from the inverse of the above coefficient matrix. The modus operandi is elaborated in more details in our previous papers \cite{Banerjee:2017gzj,Banerjee:2018hbl,Banerjee:2019lrv} \\
We were able to achieve this for our algebra and determine the supertrace elements. They are a two parameter family given by:
\begin{align}
&\left\langle J_a, P_b   \right\rangle = \eta_{ab}, \qquad 
\left\langle J_a, J_b   \right\rangle = \mu_1 \eta_{ab}, \qquad
\left\langle Q^{1,\alpha}_{\hat{p}}, Q^{1,\beta}_{\hat{q}}   \right\rangle = \left\langle Q^{2,\alpha}_{\hat{p}}, Q^{2,\beta}_{\hat{q}}   \right\rangle = 2 C_{\hat{p}\hat{q}} \eta^{\alpha \beta}, \nonumber\\ 
&\left\langle T^{\bar{a}}, Z^{\bar{b}}   \right\rangle = (12 \hat{\alpha}) \delta^{\bar a \bar b}, \qquad \qquad 
\left\langle T^{\bar{a}}, T^{\bar{b}}   \right\rangle =\mu_2 (12 \hat{\alpha}) \delta^{\bar{a}\bar{b}},
\end{align}
with $\mu_1$ and $\mu_2$ being the two arbitrary constants and they can be set to any value , even zero. As we will see in the later sections that these parameters bring interesting physics to our gravity system. 

\section{The corresponding Chern-Simons Theory and its Classical Solution}\label{sec3}
In the Chern-Simons formulation of gravity, a $2+1$ dimensional (super)gravity theory can be expressed in terms of Chern-Simons theory with a gauge group identical to that of the (super)Poincare group. The Chern-Simons is given as:

\begin{equation}\label{csaction}
S = \frac{k}{4\pi}\int_M < \mathcal{A} , d\mathcal{A} +\frac{2}{3} \mathcal{A}^2> .
\end{equation}
Here the gauge field $A$ is a Lie-algebra-valued one form and $<,>$ represents metric
in the field space that one obtains from the supertrace elements on
the Lie algebra space. $M$ is a three dimensional manifold and $k$ is the level for the CS theory. For a theory of gravity it takes the value $k=\frac{1}{4G}$ where $G$ is Newton's constant.  We express $\mathcal{A} = \mathcal{A}^a_\mu T_a dx^ \mu$ where ${T_a}$ are
a particular basis of the Lie-algebra. The equation of motion for the gauge field $A$ is given as: $F\equiv d\mathcal{A} + \mathcal{A} \wedge \mathcal{A} =0.$ 

For our supergravity theory, we consider the Lie-algebra as \eqref{eq:superPoincare} and $M$ as a three manifold with a {\it boundary}, we denote the coordinates as $u,r,\phi$. Here $u$ is the timelike coordinate and we take $r \rightarrow \infty$ hypersurface as the boundary of the space time. Thus the coordinates on boundary hypersurface are $u$ and a periodic direction $\phi$ with a period of $2 \pi$ . Expanding the gauge field in the Lie-algebra basis as

\begin{align}\label{GF}
\mathcal{A} &= e^a P_a + \hat{\omega}^a J_a + \psi^{\hat{p}}_{1,\alpha} Q^{1,\alpha}_{\hat{p}} +  + \psi^{\hat{p}}_{2,\alpha} Q^{2,\alpha}_{\hat{p}} + B^{\bar{a}} T_{\bar{a}} + C^{\bar{a}} Z_{\bar{a}}. 
\end{align}

The action \eqref{csaction} takes the following form (upto total derivatives in $d(e^a \omega_a)$ and $d(b^{\bar{a}} C_{\bar{a}})$):

\begin{align}\label{SGA}
S =\frac{k}{4\pi} \int [2e^a \hat{R_a} + \mu_1 L(\hat{\omega_a}) &- 2 \bar{\Psi}^{1,{\beta}} \nabla \Psi_{1,\beta} -2 \bar{\Psi}^{2,{\beta}} \nabla \Psi_{2,\beta}  +12\hat{\alpha}\{\mu_2 BdB -2 BdC
\nonumber \\ &+ f^{\bar{a} \bar{b}\bar{c}} (2 B_{\bar{a}} B_{\bar{b}} C_{\bar{c}} +\mu_2 B_{\bar{a}} B_{\bar{b}} B_{\bar{c}})\}].
\end{align}
This is the generalised 3D maximally $\mathcal{N}$ extended supergravity action with a gravitational anomaly term $L(\hat{\omega_a})$. In \eqref{GF},
the coefficients of each generator have their usual meaning. $e^a$ are the vielbeins and $\hat{\omega^a} = \omega^a + \gamma e^a$ with $\omega^a$ being the spin connections. $\psi_{I,\alpha}^{\hat p}$ are the gravitini fields and $B_{\hat{a}}, C_{\hat{a}}$ are the bosonic fields corresponding to automorphism. All these fields are one forms. In particular  $\psi_{I,\alpha}^{\hat p}$ are Grassmanian one-forms and $\bar{\psi}_{I,\hat{p}}^{\alpha} =  \psi^{\hat{q}}_{I,\beta}C_{\hat{q}\hat{p}}\eta^{\alpha\beta}$ . Various quantities appearing in the action \eqref{SGA} are defined as,
\begin{align*}
\hat{R_a} = d\hat{\omega_a} &+ \frac{1}{2} \epsilon_{abc}\hat{\omega^b}\hat{\omega^c}, \quad
L = \hat{\omega^a} d\hat{\omega_a} + \frac{1}{3} \epsilon^{abc} \hat{\omega_a} \hat{\omega_b}\hat{\omega_c}, \quad \\
\nabla \Psi_{1,\alpha} =& d\psi_{1,\alpha}^{\hat{p}} + \frac{1}{2} \hat{\omega}^a (\Gamma^a)^{\hat{p}}_{\hat{q}} \psi^{\hat{q}}_{1,\alpha} + B_{\bar{a}} (\lambda^{\bar{a}})^{\beta}_{\alpha} \psi^{\hat{p}}_{1,\beta}\\
\nabla \Psi_{2,\alpha} =& d\psi_{2,\alpha}^{\hat{p}} + \frac{1}{2} \hat{\omega}^a (\Gamma^a)^{\hat{p}}_{\hat{q}} \psi^{\hat{q}}_{2,\alpha} - B_{\bar{a}} (\lambda^{\bar{a}})^{\beta}_{\alpha} \psi^{\hat{p}}_{2,\beta}.
\end{align*}

The equations of motion for various fields take the following forms :
\begin{align}\label{eom}
    & de_a + \epsilon_{abc} e^b \hat{\omega}^c -\frac{1}{4} \psi^{\hat{p}}_{1,\alpha}\psi^{\hat{q}}_{1,\beta} \eta^{\alpha \beta} (C\Gamma_a)_{\hat{p}\hat{q}} -\frac{1}{4} \psi^{\hat{p}}_{1,\alpha}\psi^{\hat{q}}_{1,\beta} \eta^{\alpha \beta} (C\Gamma_a)_{\hat{p}\hat{q}}= 0 \nonumber \\
    & d \hat{\omega}_a + \frac{1}{2} \epsilon_{abc} \hat{\omega}^b \hat{\omega}^c = 0 \nonumber \\
    & d \psi^{\hat{p}}_{1,\alpha} + \frac{1}{2} {\hat{\omega}}^a \psi^{\hat{q}}_{1,\alpha} (\Gamma_a)^{\hat{p}}_{\,\,\,\hat{q}} + \psi^{\hat{p}}_{1,\beta} B^{\bar{a}} (\lambda_{\bar{a}})^{\beta}_{\,\,\,\alpha} = 0 \nonumber \\
     & d \psi^{\hat{p}}_{2,\alpha} + \frac{1}{2} {\hat{\omega}}^a \psi^{\hat{q}}_{2,\alpha} (\Gamma_a)^{\hat{p}}_{\,\,\,\hat{q}} - \psi^{\hat{p}}_{2,\beta} B^{\bar{a}} (\lambda_{\bar{a}})^{\beta}_{\,\,\,\alpha} = 0\\
     & d B_{\bar{a}} + \frac{1}{2} f_{\bar{a}\bar{b}\bar{c}} B^{\bar{b}} B^{\bar{c}} =0 \nonumber \\
     & d C_{\bar{a}} + f_{\bar{a}\bar{b}\bar{c}} B^{\bar{b}} C^{\bar{c}} + \frac{1}{12 \hat{\alpha}}(\psi^{\hat{p}}_{1,\alpha} \psi^{\hat{q}}_{1,\beta}  C_{\hat{p}\hat{q}}(\lambda_{\bar{a}})^{\alpha \beta} - \psi^{\hat{p}}_{2,\alpha} \psi^{\hat{q}}_{2,\beta}  C_{\hat{p}\hat{q}}(\lambda_{\bar{a}})^{\alpha \beta}) = 0. \nonumber 
\end{align}

We can also write down the fermion equation for its upper internal component as,
\begin{align}
 d\psi^{\hat{p},\alpha}_{1} + {\hat{\omega}}^{\hat{p}}_{\,\,\,\hat{q}}  \psi^{\hat{q},\alpha}_{1}  - \psi^{\hat{p},\gamma}_{1} B_{\gamma}^{\,\,\,\alpha} = 0 .
\end{align}

\subsection{Classical Solutions}
Even though $2+1$ dimensional (super)gravity is a topological theory, we see that the classical equations of motion do not fix the classical profile completely even after gauge fixing. There are residual gauge degrees of freedoms left unfixed at the boundary. In this section we find the classical solutions for our theory up to these unfixed residual degrees of freedoms.

We begin with equation for the $B$ field from above \eqref{eom}. Contracting the equation with $\lambda_{\bar{a}}$ and using the property $[\lambda_{\bar{a}},\lambda_{\bar{b}}] = f_{\bar{a} \bar{b}}^{\bar{c}} \lambda_{\bar{c}}$ we get:
\begin{equation}
dB + B\wedge B = 0,
\end{equation}
where we define $B^{\bar{a}} \lambda_{\bar{a}} = B$ . The solution is given as $B = A^{-1}dA$, where $A(r,u,\phi)$ is a matrix valued field in the automorphism algebra space . Similarly the equation for $\omega$ solves to $\hat{\omega} = \frac{1}{2} \hat \omega^a
\Gamma_a=\Lambda^{-1}d\Lambda$. Here $\Lambda(r,u,\phi)$ is a $SL(2,R)$ group element and $\Gamma_a$ are the generators of $SL(2,R)$. The other equations can be solved using the above two solutions of $B$ and $\omega$. Below we list the solutions of all the fields with their explicit index structures. 

\begin{align}
 e^{\hat{p}}_{\,\,\,\hat{q}} &= (\Lambda^{-1})^{\hat{p}}_{\,\,\,\hat{m}}[\frac{1}{2}(A^{\alpha\beta}\eta_{1,\beta}^{\hat{m}} A_{\alpha\sigma}d\bar{\eta}_{1,\hat{n}}^{\sigma} + \frac{1}{2} A_{\alpha\sigma}d\bar{\eta}_{1,\hat{r}}^{\sigma} A^{\alpha\beta}\eta_{1,\beta}^{\hat{r}} \delta^{\hat{m}}_{\,\,\,\hat{n}}) \nonumber\\
 &+\frac{1}{2}((A^{-1})^{\alpha\beta}\eta_{1,\beta}^{\hat{m}} (A^{-1})_{\alpha\sigma}d\bar{\eta}_{1,\hat{n}}^{\sigma} + \frac{1}{2} (A^{-1})_{\alpha\sigma}d\bar{\eta}_{1,\hat{r}}^{\sigma} (A^{-1})^{\alpha\beta}\eta_{1,\beta}^{\hat{r}} \delta^{\hat{m}}_{\,\,\,\hat{n}})+ (db)^{\hat{m}}_{\,\,\,\hat{n}}] (\Lambda)^{\hat{n}}_{\,\,\,\hat{q}}
 \nonumber \\
    \omega^{\hat{p}}_{\,\,\,\hat{q}} &= (\Lambda^{-1} d\Lambda)^{\hat{p}}_{\,\,\,\hat{q}} 
    \qquad 	B^\alpha_{\,\,\,\beta} = (A^{-1} d A)^\alpha_{\,\,\,\beta} \nonumber \\
    \quad    C_{\sigma \gamma} &= \frac{1}{2}\Big[ (A^{-1})_{\sigma\beta} \bar{\eta}_{1,\hat{q}}^{\beta} (A^{-1})_{\gamma\xi} d{\eta}^{\hat{q},\xi}_{1} - A_{\sigma\beta} \bar{\eta}_{2,\hat{q}}^{\beta} A_{\gamma\xi} d{\eta}^{\hat{q},\xi}_{2} + (d\tilde{C})_{\sigma\gamma}\Big] \nonumber \\
	\quad \psi^{\hat{p},\alpha}_{1} &= (A^{-1})^{\alpha}_{\,\,\,\beta} (\Lambda^{-1})^{\hat{p}}_{\,\,\,\hat{q}} d\eta^{\hat{q},\beta}_{1} 
		\qquad \psi_{2,\alpha}^{\hat{p}} = (A)_{\alpha}^{\,\,\,\beta}(\Lambda^{-1})^{\hat{p}}_{\,\,\,\hat{q}} d\eta_{2,\beta}^{\hat{q}}. \label{eqn:EOMsol}
\end{align}

In the above set of solutions, all fields are functions of all three coordinates $r,u,\phi$. We further impose a radial gauge condition $\partial_\phi \mathcal {A}_r=0$ to remove the redundant gauge freedom of the solutions and hence a group element splits as $G(u, \phi, r) = g(u, \phi)h(u, r)$. In this choice of  gauge, the field $\mathcal{A}$ takes following form,
$$ \mathcal {A} = h^{-1}(a + d)h, \quad  a = g^{-1}d g = a_u(u, \phi)d u + a_{\phi}(u, \phi)d \phi.$$
We choose $ h = e^{-r P_0}$ asymptotically and hence $\dot{h}(u, r) = \frac{\partial h(u,r)}
{\partial u} = 0$ at the
boundary $r \rightarrow \infty$. By this choice the dependence on the radial coordinate
is completely absorbed by the group element $h$ and the boundary can be assumed to be
uniquely located at any arbitrary fixed value of $r = r_0$, in particular to infinity. Thus the
boundary describes a two-dimensional time-like surface, described by coordinated $u,\phi$ with the topology of a cylinder. The
radial gauge choice implies that the solutions can further be decomposed into,

\begin{align}\label{decomposedsol}
   \Lambda^{\hat{p}}_{\,\,\,\hat{q}} &= (\lambda(u,\phi))^{\hat{p}}_{\,\,\,\hat{m}}(\zeta(u,r))^{\hat{m}}_{\,\,\,\,\hat{q}},~~ (\Lambda^{-1})^{\hat{p}}_{\,\,\,\hat{q}} =  (\zeta(u,r)^{-1})^{\hat{p}}_{\,\,\,\hat{m}}(\lambda(u,\phi)^{-1})^{\hat{m}}_{\,\,\,\,\hat{q}} \nonumber \\
   A^{\alpha}_{\,\,\,\beta} &=(\gamma(u,\phi))^{\alpha}_{\,\,\,\sigma}(\delta(u,r))^{\sigma}_{\,\,\,\beta},~~(A^{-1})^{\alpha}_{\,\,\,\beta} =(\delta(u,r)^{-1})^{\alpha}_{\,\,\,\sigma}(\gamma(u,\phi)^{-1})^{\sigma}_{\,\,\,\beta} \nonumber \\
   \eta_{1}^{\hat{p},\alpha} &=(\gamma(u,\phi))^{\alpha}_{\,\,\,\beta} \ [(\lambda(u,\phi))^{\hat{p}}_{\,\,\,\hat{q}}\tilde{d}_{1}^{\,\hat{q},\beta}(u,r)+d_{1}^{\,\hat{p},\beta}(u,\phi)]\nonumber \\
   \eta_{2,\alpha}^{\hat{p}} &=(\gamma(u,\phi)^{-1})_{\alpha}^{\,\,\,\beta}  [(\lambda(u,\phi))^{\hat{p}}_{\,\,\,\hat{q}}\tilde{d}_{2,\beta}^{\,\hat{q}}(u,r)+d_{2,\beta}^{\,\hat{p}}(u,\phi)]\nonumber \\
   \bar{\eta}_{1}^{\hat{p},\alpha} &=(\gamma(u,\phi))^{\alpha}_{\,\,\,\beta} [{\bar{\tilde{d}}}_{1}^{\,\hat{q},\beta}(u,r)(\lambda(u,\phi)^{-1})_{\hat{q}}^{\,\,\,\hat{p}}+{\bar{d}}_{1}^{\,\hat{p},\beta}(u,\phi)] \nonumber \\
   \bar{\eta}_{2,\alpha}^{\hat{p}} &= (\gamma(u,\phi)^{-1})_{\alpha}^{\,\,\,\beta} [{\bar{\tilde{d}}}_{2,\beta}^{\,\hat{q}}(u,r)(\lambda(u,\phi)^{-1})_{\hat{q}}^{\,\,\,\hat{p}}+{\bar{d}}_{2,\beta}^{\,\hat{p}}(u,\phi)]  \nonumber \\
   \tilde{C}_{\,\alpha,\beta} &= c(u,\phi)_{\alpha,\beta} + \tilde{c}(u,r)_{\alpha,\beta} \nonumber \\
   &- \kappa\bar{d}_{1,\hat{p},\alpha}(u,\phi) (\lambda (u,\phi))^{\hat{p}}_{\,\,\,\hat{q}} \tilde{d}_{1,\beta}^{\,\hat{q}}(u,r) + \frac{1}{\kappa}\bar{d}_{2,\hat{p},\alpha}(u,\phi) (\lambda (u,\phi))^{\hat{p}}_{\,\,\,\hat{q}} \tilde{d}_{2,\beta}^{\,\hat{q}}(u,r) \nonumber \\
   b^{\,\hat{p}}_{\,\,\,\hat{q}} &= F(u,\phi)^{\hat{p}}_{\,\,\,\hat{q}} + (\lambda(u,\phi))^{\hat{p}}_{\,\,\,\hat{m}}(E(u,r))^{\hat{m}}_{\,\,\,\hat{n}}(\lambda(u,\phi)^{-1})^{\hat{n}}_{\,\,\,\hat{q}} \nonumber \\
   &- \kappa\big[ d_{1,\alpha}^{\hat{p}}(u,\phi) (\lambda^{-1}(u,\phi))_{\hat{q}}^{\,\,\,\hat{m}} {\bar{\tilde{d}}}_{1,\hat{m}}^{\,\alpha}(u,r) + (\lambda(u,\phi)^{-1})_{\hat{r}}^{\,\,\,\hat{m}} {\bar{\tilde{d}}}_{1,\hat{m}}^{\,\alpha}(u,r)d_{1,\alpha}^{\,\,\,\hat{r}} \delta^{\hat{p}}_{\,\,\,\hat{q}}\big] \nonumber \\
   &- \frac{1}{\kappa}\big[ d_{2,\alpha}^{\hat{p}}(u,\phi) (\lambda^{-1}(u,\phi))_{\hat{q}}^{\,\,\,\hat{m}} {\bar{\tilde{d}}}_{2,\hat{m}}^{\,\alpha}(u,r) + (\lambda(u,\phi)^{-1})_{\hat{r}}^{\,\,\,\hat{m}} {\bar{\tilde{d}}}_{2,\hat{m}}^{\,\alpha}(u,r)d_{2,\alpha}^{\,\,\,\hat{r}} \delta^{\hat{p}}_{\,\,\,\hat{q}}\big].
\end{align}
The constant $\kappa$ appearing in the above equations can be found as follows : since $A$ is a matrix in the internal algebra space, it can be expanded in terms of generators $(\lambda^{\bar{a}})^{\alpha}_{\,\,\,\gamma}: \, A^{\alpha}_{\,\,\,\beta}= \sum_{\bar{b}} X_{\bar{b}} (\lambda^{\bar{b}})^{\alpha}_{\,\,\,\beta}.$ Similarly we have $(A^{-1})^{\alpha}_{\,\,\,\beta}=\sum_{\bar{a}} Y_{\bar{a}} (\lambda^{\bar{a}})^{\alpha}_{\,\,\,\beta}$. Now demanding $ (A^{-1}A)^{\alpha}_{\,\,\,\beta}=\delta^{\alpha}_{\beta}$ we get 
\begin{align}
   \sum_{\bar{a},\bar{b}} Y_{\bar{a}} X_{\bar{b}} [f^{\bar{a}\bar{b}\bar{c}}(\lambda_{\bar{c}})^{\alpha}_{\,\,\,\beta} + c_1 \nabla^{\bar{a}\bar{b}}\delta^{\alpha}_{\beta}] = \delta^{\alpha}_{\beta},
\end{align}
$c_1$ is a constant of the group and $\nabla^{\bar a \bar b}$ is the commutator of the $\lambda^{\bar{a}}$ matrices. 
To solve the above constraint we need $Y_{\bar{a}}=\kappa X_{\bar{a}}$ which implies $A^{-1} = \kappa A$, for some constant $\kappa$. 

In the above set of equations \eqref{decomposedsol}, the $u-$ derivative of all the fields $\zeta,\delta,\tilde d_{1,2}, E,\tilde c $ are set to zero value at the boundary $r \rightarrow \infty$ . Thus, at the boundary these field are fixed and have no dynamics. On the other hand $\lambda,\gamma,d_{1,2},F$ and $c$ are the residual degrees that remain unfixed by classical eoms. In the next section we present the theory that governs the dynamics of these residual fields.\\	

\section{The dual Wess-Zumino-Witten Model}\label{sec4}

In order to write the dual WZW theory at the boundary we reduce the above Chern-Simons action in its Hamiltonian form. For a well defined variational principle, actions generically require a boundary term to be added. In this case the term looks like,
\begin{equation}
\delta I_{bdy} = -\frac{k}{2\pi}\int du\wedge d\phi\wedge dr\, \partial_r\left\langle  a_{u},\delta a_{u}\right\rangle .
\end{equation}
In principle there would be another term which we omit by choosing radial gauge as stated earlier : $ \mathcal {A}(u,r,\phi) = h^{-1}(r) (a(u,\phi)+d) h(r)$. To evaluate this boundary term we first need to fix the asymptotic structure of the CS gauge field. The bosonic part of the field $\mathcal {A}$ is fixed by demanding that we have asymptotically flat metric at the boundary. The asymptotic form of the residual gauge field is:
\begin{align}\label{a}
a & =\sqrt{2} (J_1 + \mathcal{P} J_0 + \mathcal{J} P_0 -\frac{\pi}{k} \psi_{1,\alpha} Q^{1,\alpha}_{+}- \frac{\pi}{k} \psi_{2,\alpha} Q^{2,\alpha}_{+} -\frac{2\pi}{k} \mathcal{Z}_a T^{a}  -\frac{2\pi}{k} \mathcal{T}_a Z^{a})\, d\phi \nonumber \\
 &+ (\sqrt{2} P_1 + \frac{8\pi}{k} \mathcal{Z}_a Z^a +\mathcal{P}_a P^a) du.
\end{align}
The above form implies the following constraints on the asymptotic fields:
\begin{align}
&e^a_u = \omega^a_{\phi} \hspace{20pt} \omega^a_u = 0 \hspace{20pt} \psi^{I,\alpha}_u =0 \nonumber\\
&B_u = 0 \hspace{20pt} C^{\bar{a}}_u = -4B^{\bar{a}}_{\phi}.
\end{align}
The $\phi$ component of the gauge field \eqref{a} is further constrained as,
\begin{align}\label{fcc}
  \hat{\omega}_\phi^1=\sqrt{2} ~;~ \omega_\phi^2=0 ~;~ \psi_{\phi, \alpha}^{1 -} = \psi_{\phi, \alpha}^{2 -}=0 ~;~ e_\phi^1=e_\phi^2=0.
\end{align}
With these the surface term becomes,
\begin{equation}
I_{bdy} = -\frac{k}{4\pi}\int du\, d\phi [\omega^a_{\phi} \omega_{a\phi}-4B^{\bar{a}}_{\phi}B_{\bar{a}\phi}]
\end{equation}
and the total action takes the following form:
\begin{align}
I_{tot} &= \frac{k}{4\pi} \int du\, d\phi \left\langle a_{\phi},a_u \right\rangle -\frac{k}{4\pi}\int du\, d\phi [\omega^a_{\phi} \omega_{a\phi}-4B^{\bar{a}}_{\phi}B_{\bar{a}\phi}] + \frac{1}{3} \int \left\langle A, A^2 \right\rangle \nonumber\\
=& \frac{k}{4\pi} \Big[\int du\, d\phi~\big[ \omega^a_{\phi} e_{a,u} +e^a_{\phi} \omega_{a,u} + \mu_1 \omega^a_u \omega_{a,\phi} + 2C_{\hat{p}\hat{q}} \psi^{\alpha, \hat{p}}_{I,u} \psi^{\hat{q}}_{I,\alpha,\phi} \nonumber \\
& + 12\hat{\alpha} (C^{\bar{a}}_{\phi} B_{\bar{a},u} + B^{\bar{a}}_{\phi} C_{\bar{a},u} + \mu_2 B^{\bar{a}}_{\phi}B_{\bar{a},u} ) - \omega^a_{\phi} \omega_{a\phi}+4B^{\bar{a}}_{\phi}B_{\bar{a}\phi} \big] \nonumber\\
&\,\, + \frac{1}{6} \int \big[ 3 \epsilon_{abc} \omega^a \omega^b e^c + \mu_1 \epsilon_{abc}\omega^a \omega^b \omega^c + \frac{3}{2} \omega^a (C\Gamma_a)_{\hat{p}\hat{q}} \psi^{\alpha,\hat{p}}_{I} \psi^{\hat{q}}_{I,\alpha}  \nonumber\\ 
&  + \hspace{10pt}6(\lambda^{\bar{a}})^{\alpha\beta} B^{\bar{a}} C_{\hat{p}\hat{q}}(\psi^{\hat{p}}_{1,\alpha} \psi^{\hat{q}}_{1,\beta} - \psi^{\hat{p}}_{2,\alpha} \psi^{\hat{q}}_{2,\beta})\nolinebreak + 12\hat{\alpha} f_{\bar{a}\bar{b}\bar{c}}(3B^{\bar{a}}B^{\bar{b}}C^{\bar{c}} + \mu_2 B^{\bar{a}}B^{\bar{b}}B^{\bar{c}})\big] \Big].  \label{eq:totalaction}
\end{align}

 We shall evaluate the above action on the classical solutions obtained in the last section. The last two terms of \eqref{eq:totalaction} simplify as follows,\\
\begin{align*}
	6( B^{\bar{a}} (\lambda^{\bar{a}})^{\alpha\beta}C_{\hat{p}\hat{q}}(\psi^{\hat{p}}_{1,\alpha} \psi^{\hat{q}}_{1,\beta} - \psi^{\hat{p}}_{2,\alpha} \psi^{\hat{q}}_{2,\beta}) &= - 6~Tr[(A^{-1})^{\alpha}_{\,\,\,\sigma} (dA)^{\sigma}_{\,\,\,\gamma}( (A^{-1})^{\gamma}_{\,\,\,\xi}d\bar{\eta}_{1,\hat{q}}^{\xi} (A^{-1})_{\alpha}^{\,\,\,\zeta}d\eta_{1,\zeta}^{\hat{q}} \nonumber\\
	&- A^{\gamma}_{\,\,\,\xi}d\bar{\eta}_{2,\hat{q}}^{\xi} A_{\alpha}^{\,\,\,\zeta}d\eta_{2,\zeta}^{\hat{q}})]
\end{align*}
and
\begin{align*}
	12 \hat{\alpha}f_{\bar{a}\bar{b}\bar{c}} \times 3 B^{\bar{a}} B^{\bar{b}} C^{\bar{c}} 
	   =& ~ 6 \times \Big(\frac{2D}{d(d-1)}\Big) Tr[d\{(A^{-1})^{\alpha}_{\,\,\,\sigma} (dA)^{\sigma}_{\,\,\,\gamma}( (A^{-1})^{\gamma}_{\,\,\,\xi}\bar{\eta}_{1,\hat{q}}^{\xi} (A^{-1})_{\alpha}^{\,\,\,\zeta}d\eta_{1,\zeta}^{\hat{q}} \nonumber\\
	&- A^{\gamma}_{\,\,\,\xi}\bar{\eta}_{2,\hat{q}}^{\xi} A_{\alpha}^{\,\,\,\zeta}d\eta_{2,\zeta}^{\hat{q}})\}]
	  \nonumber \\
	  &+ 6 \times \Big(\frac{2D}{d(d-1)}\Big) Tr[(A^{-1})^{\alpha}_{\,\,\,\sigma} (dA)^{\sigma}_{\,\,\,\gamma}( (A^{-1})^{\gamma}_{\,\,\,\xi}d\bar{\eta}_{1,\hat{q}}^{\xi} (A^{-1})_{\alpha}^{\,\,\,\zeta}d\eta_{1,\zeta}^{\hat{q}} \nonumber\\
	&- A^{\gamma}_{\,\,\,\xi}d\bar{\eta}_{2,\hat{q}}^{\xi} A_{\alpha}^{\,\,\,\zeta}d\eta_{2,\zeta}^{\hat{q}})]. \nonumber\\
\end{align*}

Finally the on-shell action after possible simplifications evaluates to,

\begin{align}\label{osa1}
	I_{tot} =& \frac{k}{4\pi} \Bigg[ \int du d\phi \Big[2\mu_1(\Lambda^{-1}\dot{\Lambda})^{\hat{p}}_{\,\,\,\hat{m}}( \Lambda^{-1}\Lambda^\prime)^{\hat{m}}_{\,\,\,\,\hat{p}} + 2((A^{-1})^{\alpha}_{\,\,\,\beta}(\bar{\eta}_{1,\hat{p}}^{\beta})^\prime (A^{-1})_{\alpha}^{\,\,\,\sigma}\dot{\eta}^{\hat{p}}_{1,\sigma} + A^{\alpha}_{\,\,\,\beta}(\bar{\eta}_{2,\hat{p}}^{\beta})^\prime A_{\alpha}^{\,\,\,\sigma}\dot{\eta}^{\hat{p}}_{2,\sigma})\nonumber \\ &+4(\dot{\Lambda})^{\hat{p}}_{\,\,\,\hat{m}} (\Lambda^{-1})^{\hat{m}}_{\,\,\,\,\hat{q}}\Big(\frac{1}{2}((A^{-1})^{\alpha\beta}\eta^{\hat{q}}_{1,\beta} (A^{-1})_{\alpha\sigma}({\bar{\eta}}^{\sigma}_{1,\hat{p}})^\prime + \frac{1}{2} (A^{-1})_{\alpha\sigma}(\bar{\eta}^{\sigma}_{1,\hat{n}})^\prime (A^{-1})^{\alpha\beta}\eta^{\hat{n}}_{1,\beta}  \delta^{\hat{q}}_{\hat{p}})\nonumber \\ 
	&+ \frac{1}{2}(A^{\alpha\beta}\eta^{\hat{q}}_{2,\beta} A_{\alpha\sigma}({\bar{\eta}}^{\sigma}_{2,\hat{p}})^\prime + \frac{1}{2} A_{\alpha\sigma}(\bar{\eta}^{\sigma}_{2,\hat{n}})^\prime A^{\alpha\beta}\eta^{\hat{n}}_{2,\beta}  \delta^{\hat{q}}_{\hat{p}}) + (b^{\prime})^{\hat{q}}_{\,\,\,\hat{p}} \Big) - \frac{8D}{d(d-1)} \times (A^{-1}\dot{A})^{\alpha}_{\,\,\,\sigma} ({\tilde{C}}^\prime)^{\sigma}_{\,\,\,\alpha}\nonumber \\ & -2(\Lambda^{-1} \Lambda^\prime)^{\hat{p}}_{\,\,\,\hat{m}}(\Lambda^{-1} \Lambda^\prime)^{\hat{m}}_{\,\,\,\hat{p}} + 4(A^{-1}A^\prime)^{\alpha}_{\,\,\,\sigma}(A^{-1}A^\prime)^{\sigma}_{\,\,\,\alpha}  - 2\mu_2 \times \frac{2D}{d(d-1)} \times (A^{-1}A^\prime)^{\alpha}_{\,\,\,\sigma}(A^{-1}\dot{A})^{\sigma}_{\,\,\,\alpha} \Big]\nonumber \\
	& + \frac{1}{6}\int dr du d\phi   \Big[ 6 \Big(\frac{2D}{d(d-1)}-1\Big)~(A^{-1})^{\alpha}_{\,\,\,\sigma} (dA)^{\sigma}_{\,\,\,\gamma}( (A^{-1})^{\gamma}_{\,\,\,\xi}d\bar{\eta}_{1,\hat{q}}^{\xi} (A^{-1})_{\alpha}^{\,\,\,\zeta}d\eta_{1,\zeta}^{\hat{q}} - A^{\gamma}_{\,\,\,\xi}d\bar{\eta}_{2,\hat{q}}^{\xi} A_{\alpha}^{\,\,\,\zeta}d\eta_{2,\zeta}^{\hat{q}}) \nonumber \\
	& + 6 \times \Big(\frac{2D}{d(d-1)}\Big) d\{(A^{-1})^{\alpha}_{\,\,\,\sigma} (dA)^{\sigma}_{\,\,\,\gamma}((A^{-1})^{\gamma}_{\,\,\,\xi}\bar{\eta}_{1,\hat{q}}^{\xi} (A^{-1})_{\alpha}^{\,\,\,\zeta}d\eta_{1,\zeta}^{\hat{q}} - A^{\gamma}_{\,\,\,\xi}\bar{\eta}_{2,\hat{q}}^{\xi} A_{\alpha}^{\,\,\,\zeta}d\eta_{2,\zeta}^{\hat{q}})\} \Big]
	  \nonumber \\
	&+ \frac{2\mu_1}{3}\int Tr \big[(\Lambda^{-1} d\Lambda)^3 \big] - 4\mu_2 \times \frac{2D}{d(d-1)} \times \int Tr \big[(A^{-1}dA)^3 \big]. \Bigg]
\end{align}

We have used the property  $	\overline{(\Lambda^{-1})^{\hat{p}}_{\,\,\,\hat{q}} \psi^{\hat{q}}_{1,\alpha}} = \overline{\psi^{\hat{q}}_{1,\alpha }}~ \Lambda_{\hat{q}}^{\,\,\,\hat{p}}$, where $\psi^{\hat{p}}_{1,\alpha}$ is a fermion and $\Lambda^{\hat{p}}_{\,\,\,\hat{q}}$ is a $SL(2,R)$ group element.
From the index structure of the above expression, it is clear that they form a trace. So far our analysis holds true for all maximally $\mathcal{N}$ extended super-Poincar\'{e} algebras of section \ref{sec2}. As it is evident, the first four lines on the right hand side of \eqref{osa1} are two dimensional terms whereas the remaining ones are three dimensional. Among the three dimensional terms, the second one is a total derivative and hence, can be written down as a two dimensional term up to a boundary contribution, that we ignore eventually. The last two terms are genuine three dimensional terms and they signify that, in presence of arbitrary constants $\mu_1, \mu_2$, they survive. But due to their WZW structure, their variations are purely two dimensional. Only problematic term is the first one. It is a combination of two expressions which are neither total derivatives nor exact WZW types. We see that for the special case of the $so(\mathcal{N})$ internal algebra, where $d$ and $D$ are related as $D=\frac{d(d-1)}{2}$, these two expressions cancel with each other, leaving only the total derivative term. However, this simplification is not possible for generic groups, where this relation does not hold. Thus for $so(\mathcal{N})$ automorphism algebra the above action simplifies to

\begin{align}
I_{tot} =& \frac{k}{4\pi} \Bigg[ \int du d\phi~Tr \Big[2\mu_1(\Lambda^{-1}\dot{\Lambda})( \Lambda^{-1}\Lambda^\prime) + 2(A^{-1}\bar{\eta_1}^\prime A^{-1}\dot{\eta_1} + A\bar{\eta_2}^\prime A\dot{\eta_2})\nonumber \\ &+2\dot{\Lambda} \Lambda^{-1}\Big(A^{-1}\eta_{1} A^{-1}\bar{\eta}_{1}^\prime +  A^{-1}\bar{\eta}_{1}^\prime A^{-1}\eta_{1}  \text{I} + A\eta_{2} A\bar{\eta}_{2}^\prime +  A\bar{\eta}_{2}^\prime A\eta_{2}  \text{I} + b^{\prime}\Big)\nonumber \\ & -2\Lambda^{-1} \Lambda^\prime \Lambda^{-1} \Lambda^\prime + 4 A^{-1}A^\prime A^{-1}A^\prime - 2\mu_2  A^{-1}A^\prime A^{-1}\dot{A} 
	 - 4 A^{-1}\dot{A} {\tilde{C}}^\prime  \nonumber \\
	& + A^{-1} dA(A^{-1}\bar{\eta}_{1} A^{-1}d\eta_{1} - A\bar{\eta}_{2} A d\eta_{2}) \Big]
	  \nonumber \\
	&+ \frac{2\mu_1}{3}\int Tr \big[(\Lambda^{-1} d\Lambda)^3 \big] - 4\mu_2  \int Tr \big[(A^{-1}dA)^3 \big] \Bigg].
\end{align}

 Although the WZW term is a closed form, its variation being zero, the reduction to a total derivative exact form is not possible in general due to topological obstructions \cite{Barnich:2013yka,Nappi:1993ie,Bernstein:1988zd,Salomonson:1988mk,Stone:1989vg}. This is why we have the last two explicit three dimensional terms in the above action. However their variations are purely two dimensional. For the specific case superalgebra, with $so(\mathcal{N})$ automorphism every other terms of the action reduces to purely 2D terms, as we have shown here. For this particular group the constant $\kappa$  of equation \eqref{decomposedsol} is a pure phase and can be set to unity without any loss of generality\footnote{When the internal symmetry algebra is $so(\mathcal{N})$, the matrix $A$ has unit determinant, thus  $\kappa= e^{2 \pi i/d}, d=\mathcal{N}$. It can be absorbed in the fermions as overall phase.}.  In the next section, we will hence construct the Liouville like theory for this particular case. \\
~~\\
Further using the gauge decomposed forms of the solutions and neglecting total derivatives terms the above action rightly simplifies to,

\begin{align}\label{WZW}
	I_{tot} &= \frac{k}{4\pi} \bigg[ \int du d\phi ~ Tr\Big[2\mu_1\lambda^{-1}\dot{\lambda} \lambda^{-1}\lambda^\prime + 2\Big( \, \gamma \, {\bar{d}_{1}}^\prime \, \gamma \dot{d}_{1} + \gamma^{-1}\, {\bar{d}_{2}}^\prime \, \gamma^{-1}\, \dot{d}_{2}\Big) \nonumber \\ 
	&+2\dot{\lambda} \lambda^{-1}\Big(\, \gamma \, d_{1} \gamma \, \bar{d}_{1} ^\prime +  \gamma^{-1} \, d_{2} \, \gamma^{-1} \, \bar{d}_{2}^\prime  + F^{\prime} \Big) -2\lambda^{-1} \lambda^\prime \lambda^{-1} \lambda^\prime  \nonumber \\ 
	& + 4\gamma^{-1}\gamma^\prime \gamma^{-1}\gamma^\prime - 2\mu_2  \gamma^{-1}\gamma^\prime \gamma^{-1}\dot{\gamma} - \gamma^{-1} \gamma^\prime \Big( \gamma{\bar{d}}_{1}\gamma\dot{d}_{1} - \gamma^{-1}{\bar{d}}_{2} \gamma^{-1}\dot{d}_{2}\Big) \nonumber\\
	&+ \frac{1}{2}\gamma^{-1}\gamma^ \prime {\dot{\lambda}} \lambda^{-1}\Big(  \gamma d_{1} \gamma \bar{d}_{1}  + \gamma^{-1} d_{2} \gamma^{-1}\bar{d}_{2} \Big) - \gamma^{-1}\dot{\gamma} c^\prime \Big]\nonumber \\
	& +\frac{2\mu_1}{3}\int Tr \big[(\Lambda^{-1} d\Lambda)^3 \big] - 4\mu_2  \int Tr \big[(A^{-1}dA)^3 \big] \bigg] .
\end{align}

The above $1+1$ dimensional action \eqref{WZW} describes the dual theory of $2+1$ dimensional asymptotically flat supergravity theory with $so(\mathcal{N})$ as its bulk automorphism symmetry for arbitrary $\mathcal{N}$. It is a chiral WZW model with two WZW terms, that belong to $SL(2,R)$ and the internal automorphism group $so(\mathcal{N})$ of the theory. Following \cite{Barnich:2015sca} and \cite{Banerjee:2018hbl}, one can check that the boundary theory has a global and a gauge symmetry. The global symmetry corresponds to $2+2D$ types of bosonic transformations generators and $2d$ types of fermionic transformation generators. The Noether currents corresponding to these symmetry transformations are only non-trivial along the $u-$ direction and hence they directly give the corresponding Noether charges. These global symmetry charges give rise to affine extended super-Poincare symmetry, identical to the bulk symmetry of the higher dimensional supergravity theory. Furthermore using usual CFT techniques of modified Sugawara constructions, the current algebra can be used to find the infinite dimensional mode algebra. Since the boundary theory is a gauge theory one has to take the suitable gauge invariant combinations of charges for this purpose. This analysis is straightforward but algebraically tedious and we do not present the details in this work. The resultant infinite dimensional mode algebra is identical to the particular $\mathcal{N}-$extended superBMS$_3$ algebra, a suitable flat limit of SCA which is an infinite dimensional extension of two copies of  $Osp(\mathcal{N}|2,R)$. This is the asymptotic symmetry algebra of the corresponding $2+1$ dimensional supergravity theory and is given as, 
\begin{align}\label{BMS}
[\hat{\mathfrak{J}}_n,\hat{\mathfrak{J}}_m]&=(n-m)\hat{\mathfrak{J}}_{n+m}+\frac{c_J}{12}n^3\delta_{n+m,0}\;, \quad
[\hat{\mathfrak{J}}_n,\hat{\mathfrak{M}}_m]=(n-m)\hat{\mathfrak{M}}_{n+m}+\frac{c_M}{12}n^3\delta_{n+m,0} \;,
\nonumber \\
[\hat{\mathfrak{J}}_n,\psi^{(1,2),\alpha}_{m}]&=\left(\frac{n}{2}-m \right)\psi^{(1,2),\alpha}_{n+m}\;, \qquad 
[\hat{\mathfrak{J}}_n,R^a_m]=-mR^a_{n+m}\;, \qquad\quad
[\hat{\mathfrak{J}}_n,S^a_m]=-mS^a_{n+m}\;, 
\nonumber \\
[R^a_n,R^b_m]&=n\,\hat\alpha\, c_R\delta^{ab}\delta_{n+m,0}+if^{abc}R^c_{n+m}\;,\quad
[R^a_n,S^b_m]=n\,\hat\alpha\, c_M\delta^{ab}\delta_{n+m,0}+if^{abc}S^c_{n+m}\;,
\nonumber \\
[R^a_n,r^{1,\alpha}_p]&= i (\lambda^a)^{\alpha}_{\beta}r^{1,\beta}_{n+p}\;, \qquad\qquad\qquad\qquad\;\; [R^a_n,r^{2,\alpha}_p]=- i (\lambda^a)^{\alpha}_{\beta}r^{2,\beta}_{n+p} \nonumber \\
\{\psi^{1,\alpha}_n,\psi^{1,\beta}_m\}&=\frac{c_M}{6}n^2\delta_{n+m,0}\eta^{\alpha\beta}+\hat{\mathfrak{M}}_{n+m}\eta^{\alpha\beta}-\frac{i}{6\hat\alpha}(n-m)(\lambda^a)^{\alpha\beta}S^a_{n+m}\nonumber \\
&-\frac{1}{48\hat\alpha}(S^aS^a)_{n+m}\eta^{\alpha\beta}       -\frac{1}{144\hat\alpha^2}\{\lambda^a,\lambda^b\}^{\alpha\beta}\frac{1}{4}(S^aS^b)_{n+m}\;,
 \nonumber \\
\{\psi^{2,\alpha}_n,\psi^{2,\beta}_m\}&=\frac{c_M}{6}n^2\delta_{n+m,0}\eta^{\alpha\beta}+\hat{\mathfrak{M}}_{n+m}\eta^{\alpha\beta}+\frac{i}{6\hat\alpha}(n-m)(\lambda^a)^{\alpha\beta}S^a_{n+m}\;,
\nonumber \\
&-\frac{1}{48\hat\alpha}(S^aS^a)_{n+m}\eta^{\alpha\beta}   -\frac{1}{144\hat\alpha^2}\{\lambda^a,\lambda^b\}^{\alpha\beta}\frac{1}{4}(S^aS^b)_{n+m}\;.
\end{align}

In the above algebra, $\hat{\mathfrak{J}}_n$ are modes of stress tensor, $\hat{\mathfrak{M}}_n$ are modes of a spin two current, $R^a_n,S^a_n$
are modes of two independent D numbers of spin one currents and $\psi^{1,\alpha}_n,\psi^{2,\beta}_n$ are modes of two independent $d (=\mathcal{N})$ numbers of spin half fermionic currents of the $1+1$ dimensional boundary theory. It is further important to note that as described in \cite{Banerjee:2018hbl}, unlike in asymptotically flat  $\mathcal{N}=4$ Supergravity theories \cite{Banerjee:2017gzj} and $\mathcal{N}=2$ Supergravity theories \cite{Fuentealba:2017fck}, when we consider the generic choice of representation for the internal symmetries, the non liner terms in the anti-commutators survive in the asymptotic symmetry algebra. The same feature also holds for the corresponding SCA as given in \cite{Henneaux:1999ib}.

\subsection{Gauging the Chiral WZW Model}
The fields of the boundary chiral WZW model are not independent. The asymptotic boundary condition \eqref{fcc} of the bulk supergravity theory imposes following constraints on them,  
\begin{align}\label{NFCC}
   (\lambda^{-1} \lambda^\prime)^1 =\sqrt{2}, \quad
   (\lambda^{-1}\frac{\tilde{\alpha}}{2}\lambda)^1 = 0, \nonumber \\
   	(\lambda^{-1}d_1^\prime + \gamma^{-1} \gamma^\prime \lambda^{-1}d_1)^{-}_{\alpha} = 0,  \nonumber \\
	(\lambda^{-1}d_2^\prime - \gamma^\prime \gamma^{-1}  \lambda^{-1}d_2)^{-}_{\alpha} = 0, 
\end{align}
where 
\begin{align}
\tilde{\alpha}^{\hat{p}}_{\,\,\,\hat{q}} &= 2(F^\prime)^{\hat{p}}_{\,\,\,\hat{q}} + \gamma_{\alpha}^{\,\,\,\sigma} d^{\hat{p}}_{1,\sigma} \gamma^{\alpha}_{\,\,\,\delta} \bar{d}^{\delta \, \prime }_{1,\hat{q}} + (\gamma^{-1})_{\alpha}^{\,\,\,\sigma} d^{\hat{p}}_{2,\sigma} (\gamma^{-1})^{\alpha}_{\,\,\,\delta} \bar{d}^{\delta \, \prime}_{1,\hat{q}} \\ \nonumber& +(\gamma^{\prime}\gamma^{-1})^{\alpha}_{\,\,\,\beta} d^{\hat{p}}_{1,\alpha}\bar{d}^{\beta}_{1,\hat{q}}
-(\gamma^{-1}\gamma^{\prime})^{\alpha}_{\,\,\,\beta} d^{\hat{p}}_{2,\alpha}\bar{d}^{\beta}_{2,\hat{q}}. \nonumber
\end{align}

Here we have listed only the $2d+2$ number of  first class constraints. 
The above constraints \eqref{NFCC} can also be expressed in terms the global charges, although the explicit relation is not important for our analysis. It sets two charges corresponding to two bosonic transformations along $\Gamma_0$ to constant values and $2d$ number of fermionic charges   corresponding to $2d$ fermionic generators with lower positive component to zero values.We need to introduce $2d+2$ number of gauge fields to to gauge the corresponding symmetries. The details of the gauging prescription for a WZW model can be found in \cite{FEHER19921,Henneaux:1999ib,Barnich_2013,Barnich:2015sca}. Here we just present the gauged action as,
\begin{equation}\label{GFA}
    I[\lambda,F,\gamma,c,d_1,d_2,\Psi_1,\Psi_2,\hat A_{\mu},\tilde A_{\mu}]= I[\lambda,F,\gamma,c,d_1,d_2,\Psi_1,\Psi_2]+ I_{gauge},
\end{equation}
where $I[\lambda,F,\gamma,c,d_1,d_2,\Psi_1,\Psi_2]$ is the boundary chiral WZW action as in \eqref{WZW} and the local gauge fixing term $I_{gauge}$ is given as,
\begin{align}
   I_{gauge} &= \frac{k}{\pi} \int du d\phi \Big[\hat A (\lambda^{-1}\frac{\tilde{\alpha}}{2}\lambda)+\tilde A (\lambda^{-1} \lambda^\prime- \mu_M)\\ \nonumber  &+  	(\lambda^{-1}d_1^\prime + \gamma^{-1} \gamma^\prime \lambda^{-1}d_1)^{-}_{\alpha} \Psi_1^{\alpha}+	(\lambda^{-1}d_2^\prime - \gamma^\prime \gamma^{-1}  \lambda^{-1}d_2)^{-}_{\alpha}\Psi_2^{\alpha} \Big].
\end{align}
Here $\hat A, \tilde A, \Psi_{i}^{\alpha}, i=1,2$ are the $2d+2$ local gauge fields. More over $\hat A, \tilde A$ are along $\Gamma_0$ generator and $\Psi_{i, +}^{\alpha }=0$ . The constant $\mu_M= \mu \Gamma_1$ is introduced so that the charge can be set to the required constant value.

\section{The Reduced Phase Space and the Liouville-like theory}\label{sec5}

In order to find the reduced phase space description of the WZW model of \eqref{WZW} we follow the standard procedure of \cite{Barnich:2015sca,Banerjee:2019epe}.
Thus we expand the fields in the Chevalley-Serre basis of the
corresponding gauge group. This is also known as Gauss Decomposition of fields. In our case, we only know the basis generators for SL(2, R) explicitly, where as the same for the internal automorphism group it is unknown. Thus for the present case our decomposition is, 

\begin{align}
	\lambda = e^{\frac{\sigma \Gamma_1}{2}}e^{-\frac{\varphi \Gamma_2}{2}} e^{\tau \Gamma_0} , \quad
	F=-(\frac{\eta}{2}\Gamma_0 + \frac{\theta}{2}\Gamma_2 +\frac{\zeta}{2}\Gamma_1),
\end{align}
where $\sigma, \varphi, \tau, \eta, \theta $ are scalar fields and are functions of both $u,\phi$. Other fields are scalars with respect to $SL(2,R)$. The Gaussian decomposition is useful as in this decomposition one part of the 3D bulk part of the WZW model simplifies to a total derivative term as,
\begin{align}
	\frac{2}{3}Tr[(d\Lambda \Lambda^{-1})^3] = dr du d\phi~ \epsilon^{\nu\gamma\delta} \partial_\nu (e^{-\phi}~ \partial_\gamma \tau~ \partial_\delta \sigma).
\end{align}
Thus using Stokes' formula the bulk term can be reduced to a 2D term. Two product operators that are mostly used are given by,
\begin{align}
	\lambda^{-1}\lambda = \begin{bmatrix} 
		-\sigma^\prime \tau e^{-\phi} - \phi^\prime/2 &~~~~ -\sqrt{2} \sigma^\prime \tau^2 e^{-\phi} + \sqrt{2} \tau^\prime - \sqrt{2} \tau \phi^\prime \\
		\frac{\sigma^\prime}{\sqrt{2}} e^{-\phi} &~~~~ \tau \sigma^\prime e^{-\phi} + \phi^\prime/2 \\
	\end{bmatrix}
\end{align}
and
\begin{align}\label{F1}
	\dot{\lambda} \lambda^{-1} = \begin{bmatrix} 
		-\frac{\dot{\sigma}}{2}-\sigma\dot{\tau}e^{-\phi} &~~ \sqrt{2}\dot{\tau}e^{-\phi}\\
		\frac{\dot{\sigma}}{\sqrt{2}}-\frac{\sigma}{\sqrt{2}}\dot{\phi}-\frac{\sigma^2}{\sqrt{2}}\dot{\tau}e^{-\phi} &~~ \frac{\dot{\phi}}{2} + \sigma\dot{\tau}e^{-\phi} \\
	\end{bmatrix}.
\end{align}

Putting the above data, the reduced action \eqref{WZW} can be the written as,
\begin{align}
	I =& \frac{k}{4\pi} \bigg[ \int du d\phi \Big[\mu_1\varphi^\prime \dot{\varphi}-2\dot{\varphi}\theta^\prime  -4\sigma\dot{\tau}e^{-\varphi}\theta^\prime +4\dot{\tau}\zeta^\prime e^{-\varphi} +2\dot{\sigma}\eta^\prime -2\sigma\dot{\varphi}\eta^\prime -2\sigma^2\dot{\tau} e^{-\varphi}\eta^\prime -{\varphi^\prime}^2 \nonumber \\ 
	&+\Big\{ 2( \gamma ^{\alpha}_{\,\,\,\sigma}({\bar{d}_{1,\hat{p}}^{\sigma}})^\prime \gamma_{\alpha}^{\,\,\,\delta}\dot{d}_{1,\delta}^{\hat{p}} + (\gamma^{-1})^{\alpha}_{\,\,\,\sigma}({\bar{d}_{2,\hat{p}}^{\sigma}})^\prime (\gamma^{-1})_{\alpha}^{\,\,\,\delta}\dot{d}_{2,\delta}^{\hat{p}}) \nonumber \\ &+4(\dot{\lambda})^{\hat{p}}_{\,\,\,\hat{m}} (\lambda^{-1})^{\hat{m}}_{\,\,\,\hat{q}}(\frac{1}{2}  \gamma^{\alpha}_{\,\,\,\sigma} d^{\hat{q},\sigma}_{1} \gamma_{\alpha}^{\,\,\,\delta}(\bar{d}_{1,\hat{p},\delta})^\prime + \frac{1}{2} (\gamma^{-1})^{\alpha}_{\,\,\,\sigma}d^{\hat{q},\sigma}_{2} (\gamma^{-1})_{\alpha}^{\,\,\,\delta}(\bar{d}_{2,\hat{p},\delta})^\prime  + (F^{\prime})^{\hat{q}}_{\,\,\,\hat{p}})\nonumber \\ 
	& + 4(\gamma^{-1}\gamma^\prime)^{\alpha}_{\,\,\,\beta}(\gamma^{-1}\gamma^\prime)^{\beta}_{\,\,\,\alpha} - 2\mu_2  (\gamma^{-1}\gamma^\prime)^{\alpha}_{\,\,\,\beta}(\gamma^{-1}\dot{\gamma})^{\beta}_{\,\,\,\alpha}- (\gamma^{-1}\dot{\gamma})^{\alpha}_{\,\,\,\beta} (c^\prime)^{\beta}_{\,\,\,\alpha} \nonumber \\
	& - (\gamma^{-1} \gamma^\prime)^{\alpha}_{\,\,\,\beta} ( \gamma^{\beta}_{\,\,\,\sigma}{\bar{d}}^{\hat{p},\sigma}_{1}\gamma_{\alpha}^{\,\,\,\delta}\dot{d}_{1,\hat{p},\delta} -(\gamma^{-1})^{\beta}_{\,\,\,\sigma}{\bar{d}}^{\hat{p},\sigma}_{2}(\gamma^{-1})_{\alpha}^{\,\,\,\delta}\dot{d}_{2,\hat{p},\delta}) \nonumber\\
	&+ (\gamma^{-1}\gamma^ \prime)^{\alpha}_{\,\,\,\beta} ({\dot{\lambda}} \lambda^{-1})^{\hat{p}}_{\,\,\,\hat{q}}(\frac{1}{2} \gamma^{\beta}_{\,\,\,\sigma} d^{\hat{q},\sigma}_{1} \gamma_{\alpha}^{\,\,\,\delta} \bar{d}_{1,\hat{p},\delta}  + \frac{1}{2}(\gamma^{-1})^{\beta}_{\,\,\,\sigma}d^{\hat{q},\sigma}_{2} (\gamma^{-1})_{\alpha}^{\,\,\,\delta}\bar{d}_{2,\hat{p},\delta} ) \Big\} \Big]\nonumber \\
	&- 4\mu_2  \int Tr[(A^{-1}dA)^3] \bigg].
\end{align}

The explicit index structure of the above equation finally gives a trace. The first class constraints \eqref{NFCC} can be written in terms of these newly defined fields. Let us
first look at $(\lambda^{-1}\lambda^\prime)^1=\sqrt{2}$  condition. It reduces as,
\begin{align}\label{FCC1}
	(\lambda^{-1}\lambda^\prime)^1=\sqrt{2} \Rightarrow \sigma^\prime = \sqrt{2} e^\varphi.
\end{align}
Next we look at fermionic current constraints in equation ($\ref{NFCC}$). They are given as,
\begin{align}
	(\lambda^{-1}d_1^\prime + \gamma^{-1} \gamma^\prime \lambda^{-1}d_1)^{-}_{\alpha} = 0 \Rightarrow (\lambda^{-1} \gamma^{-1} {d_1^{N}}^\prime)^{-}_{\alpha} = 0 ,\nonumber \\
	(\lambda^{-1}d_2^\prime - \gamma^\prime \gamma^{-1}  \lambda^{-1}d_2)^{-}_{\alpha} = 0 \Rightarrow (\lambda^{-1} \gamma~ {d_2^{N}}^\prime)^{-}_{\alpha} = 0 .\nonumber \\
\end{align}
Redefining new fermionic parameters as $d^{\hat{p}}_{1,\alpha}=(\gamma^{-1})_{\alpha}^{\,\,\,\beta} d_{1,\beta}^{N,\hat{p}}$ and $d^{\hat{p}}_{2,\alpha} = \gamma_{\alpha}^{\,\,\,\beta} d_{2,\beta}^{N,\hat{p}}$, the two above conditions can be written in a compact form as,

\begin{align}\label{FCC23}
	(d_{1,\alpha}^{N-})^\prime = \frac{\sigma}{\sqrt{2}} (d_{1,\alpha}^{N+})^\prime ,~ (d_{2,\alpha}^{N-})^\prime = \frac{\sigma}{\sqrt{2}} (d_{2,\alpha}^{N+})^\prime .
\end{align}

Finally from the last constrain equation $(\lambda^{-1}\frac{\hat{\alpha}}{2}\lambda)^1=0$ we get,
\begin{align}\label{FCC4}
\eta^\prime \sigma^2 -2\zeta^\prime + 2\sigma\theta^\prime = 0 .
\end{align} 

Further, using the redefinition of fermions and \eqref{FCC1},\eqref{FCC23},\eqref{FCC4}, the reduced action can be written as,
\begin{align*}
I  & = \frac{k}{4\pi} \bigg[ \int du d\phi \Big[\mu_1\varphi^\prime \dot{\varphi}-2\dot{\varphi}\theta^\prime  +2\dot{\sigma}\eta^\prime -2\sigma\dot{\varphi}\eta^\prime  -{\varphi^\prime}^2 \nonumber \\ 
	& -2 e^\varphi ( (\dot{d}_{1,\alpha}^{N+})(d_{1}^{N+,\alpha}) + (\dot{d}_{2,\alpha}^{N+})(d_{2}^{N+,\alpha})) + \dot{\varphi}( (d_{1,\alpha}^{N+}) (d_{1}^{N+,\alpha})^\prime + (d_{2,\alpha}^{N+}) (d_{2}^{N+,\alpha})^\prime ) \nonumber \\ 
	& + Tr\Big\{ -2({\bar{d}}_{1}^{N})^\prime \dot{\gamma} \gamma^{-1} d_{1}^{N} -2{\bar{d}}_{1}^{N} \gamma^\prime \gamma^{-1} \dot{\gamma}\gamma^{-1} d_{1}^{N} +2{\bar{d}}_{1}^{N} \gamma^\prime \gamma^{-1} {\dot{d}}_{1}^{N}  \nonumber \\ 
	& + 2({\bar{d}}_{2}^{N})^\prime  \gamma^{-1}\dot{\gamma} d_{2}^{N} -2{\bar{d}}_{2}^{N} \gamma^{-1}\gamma^\prime \gamma^{-1}\dot{\gamma} d_{2}^{N} -2{\bar{d}}_{2}^{N}  \gamma^{-1}\gamma^\prime {\dot{d}}_{2}^{N} \nonumber \\
	& +   4\gamma^{-1}\gamma^\prime \gamma^{-1}\gamma^\prime - 2\mu_2  \gamma^{-1}\gamma^\prime \gamma^{-1}\dot{\gamma}- \gamma^{-1}\dot{\gamma} c^\prime  - \gamma^{-1} \gamma^\prime( \gamma{\bar{d}}_{1}\gamma \dot{d}_{1} -\gamma^{-1}{\bar{d}}_{2}\gamma^{-1}\dot{d}_{2}) \nonumber\\
	&+ \gamma^{-1}\gamma^ \prime{\dot{\lambda}} \lambda^{-1}(\frac{1}{2}  \gamma d_{1} \gamma \bar{d}_{1}  + \frac{1}{2} \gamma^{-1} d_{2} \gamma^{-1}\bar{d}_{2} ) \Big\}\Big] - 4\mu_2  \int Tr[(A^{-1}dA)^3] \bigg].
\end{align*}

The above equation is the reduced Hamiltonian form of the boundary theory. For further simplification let us consider the two fermion terms with explicit indices,  
\begin{align}
	I_{FF} &=(\frac{k}{4\pi}) \int du d\phi[\sqrt{2}\dot{\sigma}d_{1,\alpha}^{N+}(d_{1}^{N+,\alpha})^\prime+\sqrt{2}\dot{\sigma}d_{2,\alpha}^{N+}(d_{2}^{N+,\alpha})^\prime \nonumber \\
	&-\sqrt{2}\sigma(\bar{d}_{1,\alpha}^{N+})^\prime \dot{d}_{1}^{N+,\alpha} +2(\bar{d}_{1,\alpha}^{N+})^\prime \dot{d}_{1}^{N-,\alpha} -\sqrt{2}\sigma(\bar{d}_{2,\alpha}^{N+})^\prime \dot{d}_{2}^{N+,\alpha} +2(\bar{d}_{2,\alpha}^{N+})^\prime \dot{d}_{2}^{N-,\alpha}].
\end{align}

The last two terms of the above expression are same as 3rd and 4th terms up to total $\phi,u$ derivatives. Thus we get,

\begin{align}
	I_{FF} &=(\frac{k}{4\pi}) \int du d\phi[\sqrt{2}\dot{\sigma}d_{1,\alpha}^{N+}(d_{1}^{N+,\alpha})^\prime+\sqrt{2}\dot{\sigma}d_{2,\alpha}^{N+}(d_{2}^{N+,\alpha})^\prime \nonumber \\
	&- 2\sqrt{2}\sigma(\bar{d}_{1,\alpha}^{N+})^\prime \dot{d}_{1}^{N+,\alpha}  -2\sqrt{2}\sigma(\bar{d}_{2,\alpha}^{N+})^\prime \dot{d}_{2}^{N+,\alpha} ].
\end{align}

The above four terms can be further simplified up to total derivatives and reduction condition as,

\begin{align}
		I_{FF} &=(\frac{k}{4\pi}) \int du d\phi [-2 e^\phi ( \dot{d}_{1,\alpha}^{N+} d_{1}^{N+,\alpha} + \dot{d}_{2,\alpha}^{N+} d_{2}^{N+,\alpha})].
\end{align}

We redefine new fields as,~$\xi = 2(\theta + \sigma\eta) + ( d_{1,\alpha}^{N+}d_{1}^{N+,\alpha} +  d_{2,\alpha}^{N+}d_{2}^{N+,\alpha})$ and $\chi_{i,\alpha}= e^{\phi/2} d_{i,\alpha}^{N+}$, where $i=1,2$. Finally we can write the action as,

\begin{align}\label{FLA}
I  & = \frac{k}{4\pi} \bigg[ \int du d\phi \Big[\mu_1\varphi^\prime \dot{\varphi}-\dot{\varphi}\xi^\prime -{\varphi^\prime}^2 -2 ( \dot{\chi}_{1} \chi_{1} + \dot{\chi}_{2} \chi_{2}) \nonumber \\ 
		& + Tr\Big\{ -2({\bar{d}}_{1}^{N})^\prime \dot{\gamma} \gamma^{-1} d_{1}^{N} -2{\bar{d}}_{1}^{N} \gamma^\prime \gamma^{-1} \dot{\gamma}\gamma^{-1} d_{1}^{N} +2{\bar{d}}_{1}^{N} \gamma^\prime \gamma^{-1} {\dot{d}}_{1}^{N}  \nonumber \\ 
	& + 2({\bar{d}}_{2}^{N})^\prime  \gamma^{-1}\dot{\gamma} d_{2}^{N} -2{\bar{d}}_{2}^{N} \gamma^{-1}\gamma^\prime \gamma^{-1}\dot{\gamma} d_{2}^{N} -2{\bar{d}}_{2}^{N}  \gamma^{-1}\gamma^\prime {\dot{d}}_{2}^{N} \nonumber \\
	& +   4\gamma^{-1}\gamma^\prime \gamma^{-1}\gamma^\prime - 2\mu_2  \gamma^{-1}\gamma^\prime \gamma^{-1}\dot{\gamma}- \gamma^{-1}\dot{\gamma} c^\prime  - \gamma^{-1} \gamma^\prime( \gamma{\bar{d}}_{1}\gamma \dot{d}_{1} -\gamma^{-1}{\bar{d}}_{2}\gamma^{-1}\dot{d}_{2}) \nonumber\\
	&+ \gamma^{-1}\gamma^ \prime{\dot{\lambda}} \lambda^{-1}(\frac{1}{2}  \gamma d_{1} \gamma \bar{d}_{1}  + \frac{1}{2} \gamma^{-1} d_{2} \gamma^{-1}\bar{d}_{2} ) \Big\}\Big] - 4\mu_2  \int Tr[(A^{-1}dA)^3] \bigg].
\end{align}

In the above equation \eqref{FLA}, the ${\dot{\lambda}} \lambda^{-1}$ needs to be expressed as in \eqref{F1}. We have not put in the explicit form for simplicity. \eqref{FLA} is the flat limit of a super Liouville theory.

\section{Conclusion and Outlook}\label{sec6}
In this paper we have constructed the $1+1$ dimensional dual field theory for the three-dimensional maximally $\mathcal{N}$-extended supergravity. We have explicitly shown that the dual action can be expressed in a chiral WZW form for the special case, where the internal automorphism group of the bulk gravity superalgebra is $so(\mathcal{N})$ and its asymptotic symmetry algebra is a suitable flat limit of particular SCA, $Osp(\mathcal{N}|2;R)$. Finally we have performed the Hamiltonian reduction of the system and found that the reduced phase space description of it is given by a supersymmetric Liouville like theory in its flat limit. 

As was shown in \cite{Banerjee:2018hbl}, there are a few more $\mathcal{N}$-extended supergravity theories where the asymptotic symmetry groups are different than $Osp(\mathcal{N}|2;R)$. Our construction of the gravity dual as given in this paper does not hold for these theories. Let us briefly discuss the reasoning here. In our search for a gravity dual to asymptotically flat gravity theories, we proceed by drawing a parallel with the AdS/CFT correspondence. Thus we look for a two dimensional field theory that has the same symmetry as the asymptotic symmetry of the three dimensional bulk gravity theory. The prescription that we have followed for construction of such a boundary dual is based on the 3D gravity-Chern-Simons theory-WZW model duality. We demand that the two dimensional WZW theory must have an infinite dimensional symmetry same as the asymptotic symmetry of the bulk gravity theory. This imposes certain constraints on the WZW fields. When we take these constraints into account, we find that the corresponding WZW model has the same (super)algebra as that of the bulk (super)gravity theory as its current algebra. From this current algebra, one can construct an infinite dimensional mode algebra, where the modes are that of a dimension 2 stress tensor, and its super-partner , a spin $3/2$ operator of the WZW theory. Following a modified Sugawara prescription, these boundary operators are constructed by taking bilinears of the current algebra generators. It had been explicitly shown in \cite{Bina_1997} that construction of a spin $3/2$ operator as a bilinear form is only possible for SCA $Osp(\mathcal{N}|2;R)$. Our construction in the current paper justifies the same result.

Thus, for the other $\mathcal{N}$-extended supergravity theories of \cite{Banerjee:2018hbl}, the dual construction has to be different and the dual theory would not be a WZW model. It remains an interesting open problem to find these duals.
\\

\section*{Acknowledgements}

We would like to thank Glenn Barnich for many fruitful discussions about this project. We would also like to thank Ranveer Singh for a very useful communication during the course of this work. TN would like to thank IISER Bhopal for kind hospitality during the start of this project. AB would like to thank IISER Bhopal from where this work has been done. SB would like to thank IISER Bhopal for the financial support and the facilities they have provided during the course of this work. For TN, the work is partially supported by the
ERC Advanced Grant “High-Spin-Grav” and by FNRS-Belgium (convention IISN 4.4503.15). The work of NB is partially supported by SERB ECR grant. Finally we thank the people of India for their generous support to fundamental research.\\

\appendix
\section{Conventions and Notations}\label{App1}
We shall present our conventions for the various types of indices in this appendix. 
It is also important to note the metric in each of these vector spaces the whole algebra spans. In what follows, we denote these different spaces by representative indices of that space.\\

The metric in the $ab$ space (spanned by bosonic generators) is given by:
\begin{align*}
\eta_{ab} = 
\begin{pmatrix}
-1 & 0 & 0\\
0 & 1 & 0\\
0 & 0 & 1
\end{pmatrix}
\end{align*}

In the $\hat{p}\hat{q}$ space, the inherent metric is \footnote{The counting here starts with negative indices}
\begin{align*}
C_{\hat{p}\hat{q}} = 
\begin{pmatrix}
0 & 1\\
-1 & 0
\end{pmatrix}
\end{align*}

The range of the different kinds of indices are:
\begin{align}
	a,b &\rightarrow \{1,2,3\} \nonumber \\
	\bar{a},\bar{b} &\rightarrow \{1,2,...,D\} \nonumber \\ 
	\alpha,\beta &\rightarrow \{1,2,...,d\} \nonumber \\
	\hat{p},\hat{q} &\rightarrow \{\frac{1}{2},-\frac{1}{2}\} \nonumber
\end{align}
where $d=\mathcal{N},$ is the amount of supersymmetry and $D$ is the dimension of the internal algebra. \\
 
The three dimensional Dirac matrices satisfy the usual commutation relation: $\{\Gamma_a,\Gamma_b\}=2\eta_{ab}$.
They also satisfy following useful identities:
\begin{align}
    \Gamma_a\Gamma_b & = \epsilon_{abc}\Gamma^c + \eta_{ab} \bf{I} \\
    (\Gamma)^\alpha_\beta (\Gamma)^\gamma_\delta & = 2\delta^\alpha_\delta \delta^\gamma_\beta - \delta^\alpha_\beta \delta^\gamma_\delta 
\end{align}
The explicit form of the Dirac matrices are chosen as,\\
\begin{align}
    \Gamma_0 = \sqrt{2}
\begin{pmatrix}
0 & 1\\
0 & 0
\end{pmatrix},~~~
 \Gamma_1 = \sqrt{2}
\begin{pmatrix}
0 & 0\\
1 & 0
\end{pmatrix},~~~
 \Gamma_2 = 
\begin{pmatrix}
1 & 0\\
0 & -1
\end{pmatrix}
\end{align}
Furthermore we have ,
\begin{equation}
    C^T = -C \,,\hspace{0.8cm}
    C\Gamma C^{-1} = -\Gamma^T \,, \hspace{0.8cm}
    C_{\alpha\beta}C^{\beta}_\gamma = -\delta^{\alpha\gamma}
    \end{equation}
\subsection{Convention of Grading}
 In our work involving superalgebras, we have both Grassmann forms as well as differential forms. Both involve crucial minus signs when exchanging quantities, and there are more than one conventions in the literature on how to implement the combined sign rule, taking into account both the Grassmann and the form gradings (\cite{WinNT} and the references therein). In this paper we have used the Bernstein's convention (the super odd sign rule) which is as follows: For algebraic terms $\alpha_i$ and $\alpha_j$ with  homological degrees $n_i$ and $n_j$, and super-degrees $\sigma_i$ and $\sigma_j$ respectively, we take
 \begin{equation*}
     \alpha_i.\alpha_j = (-1)^{(n_i+\sigma_i).(n_j+\sigma_j)} \alpha_j.\alpha_i
 \end{equation*}

\section{Useful identities and relations}\label{App2}
In this appendix, we list some of the useful relations and simplifications used in our calculations. 
\subsection*{Useful relations for the internal symmetry group generators}
\begin{align}
	(\lambda^{\bar{a}})^{\alpha\beta}= -(\lambda^{\bar{a}})^{\beta\alpha}\,&, \hspace{0.5cm}
	(\lambda^{\bar{a}}\lambda^{\bar{b}})^{\alpha\beta} = (\lambda^{\bar{b}}\lambda^{\bar{a}})^{\beta\alpha} \,, \hspace{0.5cm}
	[\lambda_{\bar{a}},\lambda_{\bar{b}}] =f_{\bar{a}\bar{b}\bar{c}} \lambda^{\bar{c}}\\
	(\lambda^{\bar{a}})^{\beta\gamma}(\lambda^{\bar{a}})^\alpha_\delta + (\lambda^{\bar{a}})^{\alpha\gamma}(\lambda^{\bar{a}})^\beta_\delta &= \frac{C_\rho}{d-1}(2\eta^{\alpha\beta}\delta^{\gamma}_{\delta}-\eta^{\alpha\gamma}\delta^{\beta}_{\delta} - \eta^{\beta\gamma}\delta^{\alpha}_{\delta})\\
	(\lambda^{\bar{a}})^{\alpha\beta}(\lambda^{\bar{a}})_{\sigma\gamma} &=\frac{C_\rho}{d-1}(\delta^\alpha_\sigma \delta^\beta_\gamma - \delta^\alpha_\gamma \delta^\beta_\sigma)\\
	(\lambda^{\bar{a}})_{\alpha\beta}(\lambda^{\bar{a}})^{\beta\gamma} &= -C_\rho \delta^\gamma_\alpha \,, \hspace{0.8cm}
	\lambda^{\bar{a}}\lambda^{\bar{a}} = -C_\rho I\\
	(\lambda_{\bar{a}})^\alpha_\beta(\lambda^{\bar{b}})^\beta_\gamma(\lambda_{\bar{a}})^\gamma_\sigma&= \Big(i_\rho l^2-\frac{6}{\sigma_0}\Big)(\lambda^{\bar{b}})^\alpha_\sigma\\
	\frac{C_\rho}{(d-1)} &=3\hat{\alpha}, \qquad
	f^{\bar{a}\bar{b}\bar{c}} = -\frac{2D}{dC_\rho}Tr[\lambda^{\bar{a}}\lambda^{\bar{b}}\lambda^{\bar{c}}]\\
	Tr[\lambda^{\bar{a}}\lambda^{\bar{b}}] &= -\frac{d}{D} C_\rho \delta^{\bar{a}\bar{b}}\\
	\end{align}

\section*{Simplification of terms in the total action:}
The origin of the various terms in the simplified form of the action are as follows:
\begin{align}
 \mu_1 \omega^a_u \omega_{a,\phi}  \rightarrow & Tr[2\mu_1(\Lambda^{-1}\dot{\Lambda} \Lambda^{-1}\Lambda^\prime)] \nonumber \\
2C_{\hat{p}\hat{q}} \psi^{\hat{p}}_{I,\alpha,u} \psi^{\hat{q}}_{I,\alpha,\phi} \rightarrow & Tr[2(A^{-1}\bar{\eta_1}^\prime A^{-1}\dot{\eta_1} + A\bar{\eta_2}^\prime A\dot{\eta_2})] \nonumber \\
\frac{1}{6} 3 \epsilon_{abc} \omega^a \omega^b e^c~and~\omega^a_{\phi} e^a_u +e^a_{\phi} \omega^a_{u} \rightarrow & Tr[2\dot{\Lambda} \Lambda^{-1} (A^{-1}\eta_{1} A^{-1}\bar{\eta}_{1}^\prime +  A^{-1}\bar{\eta}_{1}^\prime A^{-1}\eta_{1}  \text{I} \nonumber \\
&+ A\eta_{2} A\bar{\eta}_{2}^\prime +  A\bar{\eta}_{2}^\prime A\eta_{2}  \text{I} + b^{\prime} )] \nonumber \\
C^{\bar{a}}_{\phi} B^{\bar{a}}_u + B^{\bar{a}}_{\phi} C^{\bar{a}}_u + \mu_2 B^{\bar{a}}_{\phi}B^{\bar{a}}_u  \rightarrow & Tr [2\mu_2  A^{-1}A^\prime A^{-1}\dot{A} 
	 + 4 A^{-1}\dot{A} {\tilde{C}}^\prime]\nonumber \\
\omega^a_{\phi} \omega_{a\phi}  \rightarrow & Tr[2(\Lambda^{-1} \Lambda^\prime \Lambda^{-1} \Lambda^\prime)] \nonumber \\
4B^{\bar{a}}_{\phi}B_{\bar{a}\phi}  \rightarrow & Tr[4(A^{-1}A^\prime A^{-1}A^\prime)] \nonumber \\
\mu_1 \epsilon_{abc}\omega^a \omega^b \omega^c \rightarrow & Tr[\mu_1 (A^{-1}dA)^3]~ (WZW~term~1) \nonumber \\
\mu_2 f_{\bar{a}\bar{b}\bar{c}}B^{\bar{a}}B^{\bar{b}}B^{\bar{c}}  \rightarrow & Tr[\mu_2(\Lambda^{-1}d\Lambda)^3]~(WZW~term~2) \nonumber \\
6(\lambda^{\bar{a}})^{\alpha\beta} B^{\bar{a}} C_{\hat{p}\hat{q}}(\psi^{\hat{p}}_{1,\alpha} \psi^{\hat{q}}_{1,\beta} - \psi^{\hat{p}}_{2,\alpha} \psi^{\hat{q}}_{2,\beta})~and~ B^{\bar{a}} C^{\bar{b}} B^{\bar{c}} \rightarrow & Tr[ A^{-1} dA(A^{-1}\bar{\eta}_{1} A^{-1}d\eta_{1} - A\bar{\eta}_{2} A d\eta_{2})] \nonumber
\end{align}

\subsection*{ Terms in Hamiltonian Reduced Action
}
\begin{align}
	Tr[2\mu_1(\lambda^{-1}\dot{\lambda}\lambda^{-1}\lambda^\prime)] &=  \mu \varphi^\prime \dot{\varphi} + 2\mu(\dot{\sigma}\tau^\prime e^{-\varphi}) + 2\mu(\sigma^\prime\dot{\tau} e^{-\varphi}) \nonumber\\
	Tr[4\dot{\lambda}\lambda^{-1}F^\prime] &= 4[-\frac{\dot{\varphi} \theta^\prime}{2}+\sigma\zeta^\prime e^{-\varphi}+\frac{\dot{\sigma}\eta^\prime}{2} -\frac{\sigma\dot{\varphi}\eta^\prime}{2}-\frac{\sigma^2\dot{\tau}e^{-\varphi}\eta^\prime}{2}] \nonumber\\
	Tr[-2(\lambda^{-1}\lambda^\prime)^2] &= -2(2\sigma^\prime\tau^\prime e^{-\varphi}+\frac{{\varphi^\prime}^2}{2}) \nonumber\\
	\frac{2\mu}{3} \int [(d\Lambda \Lambda^{-1})^3] &= \int du d\phi (-2\mu(\dot{\sigma}\tau^\prime e^{-\varphi}) + 2\mu(\sigma^\prime \dot{\tau} e^{-\varphi})) \nonumber\\
	Tr[2( \gamma ({\bar{d}}_{1})^\prime \gamma \dot{d}_{1} + \gamma^{-1} ({\bar{d}}_{2})^\prime \gamma^{-1} \dot{d}_{2})] &= Tr[2((\bar{d}_{1}^N)^\prime \dot{d}_{1}^N + (\bar{d}_{2}^N)^\prime \dot{d}_{2}^N) \nonumber\\
	& +  2(-({\bar{d_1}}^N)^\prime \dot{\gamma} \gamma^\prime d_1^N -\bar{d_1}^N \gamma^\prime \gamma^{-1} \dot{\gamma}\gamma^{-1} d_1^N +\bar{d_1}^N \gamma^\prime \gamma^{-1} \dot{d_1}^N)  \nonumber \nonumber\\ 
	& +  2((\bar{d_2}^N)^\prime  \gamma^\prime \dot{\gamma} d_2^N -\bar{d_2}^N \gamma^{-1} \gamma^\prime \gamma^{-1} \dot{\gamma} d_2^N -\bar{d_2}^N \gamma^{-1} \gamma^\prime  \dot{d_2}^N) ] \nonumber\\
	Tr[2\dot{\lambda} \lambda^{-1} \gamma d_{1} \gamma (\bar{d}_{1})^\prime] &= \sqrt{2} \dot{\sigma} d_{1,\alpha}^{N+}(d_{1}^{N+,\alpha})^\prime + \dot{\varphi} d_{1,\alpha}^{N+}(d_{1}^{N+,\alpha})^\prime \nonumber\\	
	Tr[2\dot{\lambda}\lambda^{-1} \gamma^{-1}  d_{2} \gamma^{-1}(\bar{d}_{2})^\prime] &=  \sqrt{2} \dot{\sigma} d_{2,\alpha}^{N+}(d_{2}^{N+,\alpha})^\prime + \dot{\varphi} d_{2,\alpha}^{N+}(d_{2}^{N+,\alpha})^\prime\nonumber\\
		\end{align}

\bibliographystyle{jhep}
\bibliography{bms3.bib}

\end{document}